\documentstyle[12pt]{article}

\begin{document}
\begin{titlepage}
\begin{flushright}
DO-TH-93/19EA \\
February 1994 \\
\end{flushright}
\vspace{20mm}
\begin{center}
{\Large
Quantum Fluctuations of the electroweak
sphaleron: Erratum and Addendum}
\vspace{10mm}

{\large  J. Baacke \footnote{e-mail: baacke@het.physik.
uni-dortmund.de}
and S. Junker} \\
\vspace{15mm}

{\large Institut f\"ur Physik, Universit\"at Dortmund} \\
{\large D-44221 Dortmund 50} \\ {\large Germany}
\vspace{40mm}
\end{center}
\end{titlepage}
We have presented recently \cite{BaaJu} an exact computation of the
fluctuation determinant of the electroweak sphaleron.
 The results disagreed substantially from
those of an earlier evaluation of this quantity \cite{CarLi}.
Moreover none of the two exact computations agreed with analytical
estimates \cite{Sha,CarMcLe} that are expected to be good at small Higgs
masses, essentially approximations in which higher
gradient terms are neglected, a fact that has been repeatedly
criticized (see e.g. \cite{HeKrSch}).

While we thought that possibly the gradient type
expansions were to blame for this discrepancy, the failure
of compatibility between approximations and exact results
can be traced back to our treatment of the tadpole
contributions. Indeed we removed all tadpole contributions
to the Higgs field {\it completely}, in a misinterpretation
of the renormalization and rescaling prescription of
Refs.  \cite{CarLi,CarMcLe}; however, a finite piece
has to be restituted. In order to understand this point
which has considerable numerical consequences we review
shortly these contributions (see \cite{DoJa}, especially
Appendix C). This cannot be done consistently within the
three dimensional asymptotic theory since the approximation
of dismissing all but the lowest Matsubara frequency is not
justified for a these divergent contributions.

The tadpole contribution including all Matsubara frequencies reads
\begin{equation}
{\cal V}[\Phi]_{Tad} =
\frac{T}{2} \sum_i       \int\frac{d^3p}{(2\pi)^3}\sum_{n=-\infty}^\infty
\frac{1}{p^2+m_i^2+(2\pi n T)^2} c_i\int d^3x (H_0^2(\vec x)-1)
\end{equation}
Here $m_i$ are the masses circulating in the loop, $c_i$ are
their couplings to the Higgs field and $H_0(\vec x)$ is the
Higgs profile. The coefficients $c_i$ can be identified as
coefficients of the terms
proportional to $H_0^2-1$ in the diagonal elements of the potential
given in Appendix A of \cite{BaaJu}, they are given below.
The momentum integral including the factor $T/2$ can be rewritten
\cite{DoJa}
\begin{equation}
\int \frac{d^3p}{(2\pi)^3} \left( \frac{1}{4 \sqrt{p^2+m^2}}
+\frac{1}{2 \sqrt{p^2+m^2}} \frac{1}{\exp{\sqrt{p^2+m^2}/T}-1}
\right)
\end{equation}
The first term is the $T=0$ contribution which goes into the
Higgs mass renormalization. The second term can be expanded at
high temperature as
\begin{equation}
\frac{T}{  \pi}\int x^2dx \frac{1}{\sqrt{x^2+m^2/T^2}}
\frac{1}{\exp{\sqrt{x^2+m^2/T^2}}-1} \approx
\frac{T^2}{24} - \frac{mT}{8\pi}
\end{equation}
up to terms of order $\ln T$ or lower.
The term quadratic in $T$ can be absorbed \cite{CarLi,CarMcLe}
into the $T$ dependence of the
vacuum expectation value of the Higgs field. The linear term is
part of the well known $T\Phi^3$ term of the effective potential
and without this contribution the latter is incomplete (see
e.g. the discussion of this term in Appendix A of \cite{HeKrSch}).
This is
the reason why our data must fail to approach the estimate based
on the effective potential.

Specifying the contributions of the different fields by \cite{BaaJu}
$m_i=M_W , c_i=M_W^2$ for altogether six components of gauge boson
and subtracted ghost fields, $m_i=M_W , c_i= (M_H^2+2M_W^2)/2$
for the three Goldstone boson fields and by $m_i=M_H , c_i=3
M_H^2/2$ for the physical Higgs field we find a contribution to the
effective action proportional to $T^2$
\begin{equation}
\frac{T^2}{8}(3M_W^2+M_H^2)\int d^3x (H_0^2-1)
\end{equation}
in agreement with standard results. The term linear in $T$ yields
now
\begin{equation}
-\frac{T}{8\pi}\left(M_W \left(9M_W^2+\frac{3}{2}M_H^2\right)
+\frac{3}{2}M_H^3\right)\int d^3x (H_0^2-1).
\end{equation}
This term is positive. Since the fluctuation determinant is related to
\\$-{\cal V}_{1-loop}/T$ and the term is to be restituted in order
to parallel the treatment of the effective potential, we find
a large negative contribution to $\ln \kappa$.

We present in Table 1 our previous results with the
completely removed tadpoles as $\ln \tilde{\kappa}$,
in units of $(gv)^6$ as
used in \cite{CarLi} and the results with the restituted piece as
$\ln \kappa$. It is the latter one that is correct and that has
to be compared with the
$\Phi^3$ estimate derived from the effective potential which,
if only the gauge loops are taken into account, takes the form
\cite{Sha}
\begin{equation}
\frac{3M_W^3}{4\pi}\int d^3x (H_0^3-1)
\end{equation}
We give the corresponding estimates in the last row of Table 1,
labelled ``$\Phi^3$''. This estimate differs from the one given
in \cite{CarLi} by a factor $\sqrt{8}$ which is due to a
mistake there: the
term is originally \cite{Sha} given for a Higgs
field normalized to the vacuum
expectation value $v$, while the one used in \cite{CarLi} has
vacuum expectation value $v/\sqrt{2}$ \footnote{This statement is
made with the agreement of Larry McLerran}.
We observe that our corrected data and this estimate are now
well consistent. Actually, since $\kappa$ is a dimensionful
quantity $\ln \kappa$ depends on the scale used, therefore
 one can only expect
that the exact data and the estimate become {it parallel} as $M_H \to 0$;
the absolute agreement in this limit is somewhat fortitious.

Our corrected data, the analytic estimate and the data of Carson
et al. \cite{CarLi} are presented in Figure 1, showing now a reasonable
general agreement. Whether the remaining differences are within or
outside the error margins of the involved numerical computations is
a question that seems hard to answer. Unfortunately also the disagreement
with the estimate of Ref. \cite{CarMcLe} persists.

\noindent
{\bf Acknowledgments}

One of us (J.B.) thanks K. Goeke and A. Ringwald for making him aware,
at different times, of Ref. \cite{HeKrSch}, W. Buchmueller and
L. McLerran for most useful discussions, and the DESY
for hospitality.

\vspace{20mm}
\begin{center}
\section*{Table 1}
\vspace{6mm}
\begin{tabular}{|c||c|c|c|c|c|c|c|}\hline
$\xi$ & 0.4 & 0.5 & 0.6 & 0.8 & 1.0 & 1.5 & 2.0 \\  \hline
$\lambda/g^2$ &.02&.031&.045&.08&.125&.281&.5\\ \hline
$\ln \tilde{\kappa}$ &2.02&1.73&1.52&1.34&1.32&1.46&1.55\\ \hline
$\ln \kappa$ &-46.80&-31.00&-22.30&-13.64&-9.64&-5.96&-5.14\\ \hline
$\Phi^3$ &-45.23&-29.43&-20.68&-11.87&-7.73&-3.57&-2.08\\ \hline
\end{tabular}
\end{center}
\section*{Figure Captions}

\noindent
Fig. 1 {\bf The Fluctuation Determinant}

We plot the logarithm of the fluctuation determinant $\kappa$
as a function of the ratio $\lambda/g^2$. Our corrected results
are given as triangles, those of Ref. \cite{CarLi} as squares.
The solid line is the estimate based on the effective potential.
\end{document}